# Trajectory end point distribution of a test particle in a turbulent liquid

S. F. Edwards ,
Cavendish Lab. Madingley Rd.
CB30HE Cambridge, UK

Moshe Schwartz,
Raymond and Beverly Sackler School
of Physics and Astronomy
Tel Aviv University,   Tel Aviv 69978, Israel

In a recent paper the mean square displacement (MSD), $\langle R^2(T) \rangle$, of a particle carried by a turbulent liquid over time $T$ has been shown to be proportional to $T^{6/5}$ , meaning that the motion of the particle is slightly super-diffusive. In some cases, though, it might be important to have more information than that provided by that law. An example would be the distribution of pollutants as a function of time by turbulent flow. Here small amounts of material reaching relatively large distances are of importance. This motivates our interest in the full distribution of the location of particles swept by the fluid as a function of time. The distribution depends on the distance through the dimensionless quantity $X^2 = R^2 / \langle R^2(T) \rangle$. We find that for small $X$ , the distribution $P(R,T)$ is proportional to $\exp(-aX^2)$ but at its tail when $X$ is large it behaves as $\exp(-bX^{2/3})$.

pacs numbers-02.50.Ey, 05.40.Fb

The Richardson 1926 paper [1] on the relative diffusion of two particles in the atmosphere can be identified probably as the single most influential experimental work in the field of turbulence. His observation that the mean square distance between two balloons in the atmosphere is proportional to $T^3$, the cube of the elapsed time since their relies, influenced the work of Kolmogorov [2,3] and Obukhov [4] in the early forties of the 20[th] century as well as the more modern field of passive scalar turbulence [5,6]. Clearly, relative diffusion is important since it characterizes the spread of a smoke plume carried by turbulent flow and might be thus relevant to the spread of pollutants. It is clear though, that the mean square displacement of a typical particle of the plume, relative to the point where it is emitted, may be not less relevant for assessing the risk of pollutants spread. For many years this has been considered to be not a very interesting question, since an old argument of G. I. Taylor [7], states that a single particle diffuses on large time scales, while at shorter times it might be ballistic. The Taylor argument is based on the assumption that the velocities of a particle swept by a turbulent liquid, have a finite correlation time, $\tau_C$. The effective diffusion constant, describing the long time diffusion of the particle, is given by

$$D = \langle \mathbf{v}^2 \rangle \tau_C \ , \tag{1}$$

where $\mathbf{v}$ is the velocity of the particle, $\langle \cdots \rangle$ denotes an average over all possible trajectories, and the frame of reference is chosen to render the average velocity zero. The only possible weakness in the Taylor argument is the assumption that the correlation time is finite. Indeed, various problems considered over the years are characterized by an infinite correlation time, which produces super-diffusive behavior [8-11]. Namely the mean square distance (MSD) increases more than linearly with the elapsed time. In a recent paper [12] the problem of the behavior of the MSD has been revisited. It has been found, that the Fourier components of the velocity relevant to the large time scale motion of the particle are components with very small wave vectors, which are within the range of wave vectors where energy is pumped into the liquid. Consequently, the inertial range correlations (Kolmogorov) are not relevant to the determination of long time MSD, which is determined by the energy pumping range correlations and found to be given by

$$\langle R^2(T) \rangle \propto T^{6/5} \ . \tag{2}$$

Namely, we obtain slight super diffusion in contrast with simple diffusion that follows from the simple Taylor argument given above,

which assumes a finite correlation time of the Lagrangian velocities of the particle.

The above relation tells us something about the way a representative particle in a dust cloud moves in time due to turbulent air. If that dust cloud is a dangerous pollutant, for example, we would like to have much more detailed information than that given by equation (2). Such information is encoded in the probability distribution of the displacement $\mathbf{R}$ of the particle as a function of $T, P(\mathbf{R},T)$. The mean square displacement (MSD) given by eq. (2), is only one of many parameters characterizing the distribution,

$$\langle R^2(T) \rangle = \int d\mathbf{R} \, R^2 P(\mathbf{R},T) = aT^{6/5} . \qquad (3)$$

The purpose of the present article is to beyond the MSD and give the form of the displacement distribution in two extreme cases
$R^2 / \langle R^2(T) \rangle \ll 1$ and $R^2 / \langle R^2(T) \rangle \gg 1$ .

The equation describing the motion of a test particle swept by a given specified velocity field, $\mathbf{V}(\mathbf{r},t)$, is given by

$$\dot{\mathbf{R}} = \mathbf{V}(\mathbf{R}(t),t) . \qquad (4)$$

The actual velocity field is assumed to be random with zero mean and given correlations,

$$\langle \mathbf{V}_i(\mathbf{r}_1,t_1) \mathbf{V}_j(\mathbf{r}_2,t_2) \rangle = \Phi_{ij}(\mathbf{r}_2 - \mathbf{r}_1, t_2 - t_1) \equiv \Phi_{ij}(\mathbf{r}_{12},t_{12}) , \qquad (5)$$

where $i$ and $j$ denote Cartesian components.

We will proceed at present with the general form above and introduce the correlations specific to a turbulent fluid later.

Our first step is to obtain the probability distribution for finding a particular trajectory , $\mathbf{R}(t)$, in the space of trajectories connecting the origin at time $t=0$ and $\mathbf{R}$ at $t=T$ .

We will obtain now the trajectory end point distribution from the velocity correlations given by equation (5). We will assume that the velocity distribution is a Gaussian, defined by the correlations given above.

$$\Gamma\{\mathbf{V}(\mathbf{r},\mathbf{t})\} \propto \exp\{-\frac{1}{2}\int_0^T\int_0^T dt_1 dt_2 \int\int d\mathbf{r}_1 d\mathbf{r}_2 \mathbf{V}_i(\mathbf{r}_1,t_1)[\Phi^{-1}]_{ij}\mathbf{V}_j(\mathbf{r}_2,t_2)\} \quad (6)$$

where the spatial integrations are over the volume of the liquid and $\Phi^{-1}$ is the matrix inverse of the velocity correlations (5). (Namely, $\sum_k \int_0^T dt' \int d\mathbf{r}' \Phi_{ik}(\mathbf{r}_2-\mathbf{r}',t_2-t')[\Phi^{-1}]_{kj}(\mathbf{r}',\mathbf{r}_1;t',t_1) = \delta_{ij}\delta(\mathbf{r}_2-\mathbf{r}_1)\delta(t_2-t_1)$.)

Clearly, the velocity distribution in a turbulent fluid is not Gaussian. Such distributions have been used, however, in the literature to model a turbulent liquid, even with the extreme $\delta$ function correlation in time [13]. The reason is that in spite its relative simplicity, it still leads to rich and interesting behavior [13,14]. Furthermore, in the region relevant to our discussion, which corresponds to extreme space and time separation, the Gaussian distribution is a good lowest order approximation of the real distribution [15].

The probability of a given trajectory, $\mathbf{R}(t)$ is given by [14,16,17]

$$P\{\mathbf{R}(T)\} = \left\langle \prod_t \delta(\dot{\mathbf{R}} - \mathbf{V}(\mathbf{R},t)) \right\rangle , \quad (7)$$

where the average is over the distribution of the velocity field. Introducing the standard representation of the $\delta$ function and a vector function, $\mathbf{k}(t)$, we write

$$P\{\mathbf{R}(t)\} \propto \left\langle \int D\mathbf{k}(t) \exp i \int_0^T dt \mathbf{k}(t)(\dot{\mathbf{R}}(t) - \mathbf{V}(\mathbf{R}(t),t)) \right\rangle \quad (8)$$

where $D$ denotes path integration. Averaging over the velocity distribution we obtain

$$P\{\mathbf{R}(t)\} \propto \int D\mathbf{k}(t) \exp i \int_0^T dt \mathbf{k}(t) \cdot \dot{\mathbf{R}}(t)$$
$$\exp[-\frac{1}{6}\int_0^T\int_0^T dt_1 dt_2 \Phi_{ij}(\mathbf{R}(t_2)-\mathbf{R}(t_1),t_2-t_1)\mathbf{k}_i(t_1)\mathbf{k}_j(t_2)]. \quad (9)$$

We are interested in the long time statistics of the motion of the test particle. In such times the particle has already traversed long distances. Consequently, only the Fourier components of the velocity field with very small wave vectors are relevant to the motion of the particle. Such wave vectors are within the momentum range over which energy is pumped into the liquid by an external driving noise. (For a detailed discussion of this point, see ref. [12].) The Fourier transform of the velocity correlation of an incompressible turbulent liquid, for wave vectors in the energy pumping range has the form [12,15] ,

$$\langle V_i(\mathbf{q},t'+t)V_j(\mathbf{p},t')\rangle = [\delta_{ij} - \frac{q_i q_j}{\mathbf{q}^2}]Aq^{-5/3}\exp(-Bqt^{3/5})\delta(\mathbf{q}+\mathbf{p}) \ . \tag{10}$$

( For further discussion of the stretched exponential form of the scaling function see also [18-19]). The form above implies that the velocity correlation has the form

$$\Phi_{ij}(\mathbf{r}_{12},t_{12}) = r_{12}^{-4/3}[\delta_{ij} + c[\mathbf{r}_{12}^i \mathbf{r}_{12}^j/\mathbf{r}^2]]g(Bt_{12}^{3/5}/r_{12}) \ , \tag{11}$$

where $c$ is a dimensionless constant and $\mathbf{r}_{12}^i$ is a Cartesian component of $\mathbf{r}_{12}$. The function $g(x)$ is a non vanishing constant at small $x$ and at large $x$ it behaves as $x^{-4/3}$.
 The difficulty now is to perform the functional integration over $\mathbf{k}(t)$. The approximation used in ref. [12] , which enables that integration, is to replace $\Phi_{ij}$ by its average over trajectories. This involves replacing $|\mathbf{r}_{12}|$ by $|t_{12}|^\nu$, where $\nu$ is the exponent describing the MSD at large time separations, $\langle \mathbf{r}_{12}^2 \rangle \propto |t_{12}|^\nu$ .When this is done self consistently $\nu$ is found to be 3/5 ,so that the MSD scales as $\langle R^2(T)\rangle \propto T^{6/5}$ [12] .

The required trajectory end point distribution is given by

$$P(\mathbf{R},T) = \int D\mathbf{R}(t)P\{\mathbf{R}(t)\} . \tag{12}$$

(Recall that the trajectories connect (0,0) to $(\mathbf{R},T)$. The distribution is expected to have the form

$$P(\mathbf{R},T) = \frac{1}{\langle R^2(T)\rangle^{3/2}} f(|\mathbf{R}|/\langle R^2(T)\rangle^{1/2}) . \tag{13}$$

It is rather clear that when the argument of $f$ is small, $f$ is Gaussian. This can be seen in many ways. For example it is possible to expand the logarithm of $f$ to first order in its argument or it is possible to view $P(\mathbf{R},T)$ as the average of $\delta[\mathbf{R}(T)-\mathbf{R}]$ over all trajectories emanating from the origin at time 0 and then the average is performed by expanding the $\delta$ function in plane waves and averaging each of the plane waves, using the cummulant expansion to lowest order.

When $\mathbf{R}^2/\langle R^2(T)\rangle \gg 1$ the situation is more interesting. The contribution to the path integral (12) yielding the end point distribution function, comes from rare trajectories, which are relatively stretched. In the following we will present an approximation which takes that stretching into account in the crudest way and yields the behavior at the tail of the trajectory end point distribution. We have obtained a more refined but also more complicated derivation, which will not be presented here because of space limitations. Both derivations yield essentially the same result. Recall that previously we have replaced in the velocity field correlation function the distance $r_{12}$ by some dimensional constant times the time difference to the unknown power, $\nu$ and that led to a self consistent equation, which enables the determination of $\nu = 3/5$. The approximation we employ for the stretched trajectories is that at least the dominant among them are uniformly stretched. This implies that within that set of rare trajectories the replacement of $r_{12}$ by a constant times a power of $t_{12}$ has to be done in a way that takes that stretching into account. Namely,

$$r_{12} = (t_{12}/T)^{3/5} R, \qquad (14)$$

instead of

$$r_{12} = C t_{12}^{3/5}. \qquad (15)$$

The functional integration over $\mathbf{k}(t)$ can still be performed to obtain the probability distribution for a trajectory, $P\{\mathbf{R}(t)\}$, which is still Gaussian. The important new feature in that Gaussian is that it depends explicitly on the parameter $R/CT^{3/5}$. Because $P\{R(t)\}$ is Gaussian the trajectory end point distribution can be calculated explicitly but the result is **not** a Gaussian because of the parametric dependence of $P\{\mathbf{R}(T)\}$ on $R/CT^{3/5}$. The final result is

$$P(\mathbf{R},T) \propto \exp(-\frac{R^2 T^{4/5}}{2C\langle R^2(T)\rangle R^{4/3}}) = \exp(-\lambda R^{2/3}/\langle R^2(T)\rangle^{1/3}), \qquad (16)$$

where $\lambda$ is a dimensionless constant. (For expressions of a similar form obtained in other cases of super diffusion see refs.[20-21].) Note that the tail of the distribution decays much slower than a Gaussian. Since, this is only the tail, the MSD is still determined to a very good approximation from the Gaussian approximation to the trajectory end point distribution function. Note also that the only difference between the compressed and the stretched situation arises due to the difference in the behavior of $g(x)$ for small and large values of $x$. In fact, within the same approximation, we could interpolate, assuming we know the form of the scaling function, $g$.

What we have to do is to replace the argument of $g$ in equation (11) by $CT^{3/5}/R$. Consequently the interpolated expression reads

$$P(R,T) \propto \exp\{-\alpha g[\beta\langle R^2(T)\rangle^{1/2}/R](R/\langle R^2(T)\rangle^{1/2})^{2/3}, \qquad (17)$$

where $\alpha$ and $\beta$ are dimensionless constants.

The derivation above

follows from the specific form of the velocity field correlations( eq. (10)) The same procedure can be easily generalized, however, to general dimension $d$ and to more general velocity correlations of the type $A'[\delta_{ij} - \frac{\mathbf{q}_i \mathbf{q}_j}{\mathbf{q}^2}]q^{=\Gamma}\phi(B'qt^{1/z})\delta(\mathbf{q}+\mathbf{p})$, instead of the right hand side of equation (10). Consequently, the same method can be applied to the study of end point distribution in a whole range of physical systems characterized by super diffusion.
We hope to come back to the more refined derivation in the near future.

References

1. L. F. Richardson, Proc. Roy. Soc. A **110**, 709 (1926).
2. A. N. Kolmogorov, C.R. Acad. Sci. USSR **30**, 301 (1941).
3. A. N. Kolmogorov, C.R. Acad. Sci. USSR **32,** 16 (1941).